\documentclass[water,article,accept,oneauthor]{Definitions/mdpi}

\firstpage{1}
\pubvolume{1}
\issuenum{1}
\articlenumber{0}
\pubyear{2023}
\copyrightyear{2023}
\externaleditor{Academic Editor: Roberto Gaudio}
\datereceived{14 August 2023}
\daterevised{6 September 2023} 
\dateaccepted{7 September 2023}
\datepublished{ }
\hreflink{https://doi.org/} 

\Title{An Algebraic Non-Equilibrium Turbulence Model of the High Reynolds Number Transition Region}

\TitleCitation{An Algebraic Non-Equilibrium Turbulence Model of the High Reynolds Number Transition Region}

\Author{Nils T. Basse 
 \orcidA{}}

\AuthorNames{Nils T. Basse}

\AuthorCitation{Basse, N.T.}

\address[1]{Independent Researcher
, Trubadurens v\"ag 8, 423 41 Torslanda, Sweden; nils.basse@npb.dk\\
}

\abstract{We present a mixing length-based algebraic turbulence model calibrated to pipe flow; the main purpose of the model is to capture the increasing turbulence production-to-dissipation ratio observed in connection with the high Reynolds number transition region. The model includes the mixing length description by Gersten and Herwig, which takes the observed variation of the von K\'arm\'an number with Reynolds number into account. Pipe wall roughness effects are included in the model. Results are presented for area-averaged (integral) quantities, which can be used both as a self-contained model and as initial inlet boundary conditions for computational fluid dynamics simulations.}


\begin{document}


\section{Introduction}

The impetus for this paper came from an invited presentation on boundary conditions for turbulence modelling \cite{bib7}. Based on an observed variation of the turbulence production-to-dissipation ratio with Reynolds number, a corresponding variation of the turbulence model constant $C_{\mu}$ was discussed. Previously, Princeton Superpipe measurements \cite{bib18,bib36} of streamwise mean and fluctuating velocities have been analysed in detail \cite{bib5,bib6}.

We present a new algebraic (zero-equation) turbulence model based on the Prandtl mixing length concept \cite{bib29,bib30} to improve our understanding of the observed high Reynolds number transition region. It is important to note that this transition is not related to the laminar--turbulent transition, which takes place at much lower Reynolds numbers. The model can be used both standalone and to provide the initialisation of inlet boundary conditions for computational fluid dynamics (CFD) simulations \cite{bib38,bib14}. Our approach is similar to the “LIKE” algorithm in \cite{bib32}, where the letters in the abbreviation represent the integral turbulent {\bf L}
ength scale, turbulence {\bf I}ntensity, turbulent {\bf K}inetic energy and turbulent dissipation rate $\varepsilon$ ({\bf E}).

The paper is organised as follows: In Section \ref{sec:recap}, we summarise previous relevant findings, followed by a model overview in Section \ref{sec:model_overview}. Model details can be found in the Supplementary Information (SI) document \cite{bib8}. Results from our new model are provided in Section \ref{sec:mod_results} and discussed in Section \ref{sec:disc}. We conclude and propose future research directions in Section \ref{sec:conc}.

\section{The High Reynolds Number Transition Region}
\label{sec:recap}

In this section, we summarise earlier findings which prompted this study.

We begin by defining the friction Reynolds number:

\begin{equation}
Re_{\tau} = \frac{u_{\tau} \delta}{\nu_{\rm kin}},
\end{equation}

\noindent where $u_{\tau}$ is the friction velocity, $\delta$ is the boundary layer thickness (pipe radius $R$ for pipe flow) and $\nu_{\rm kin}$ is the kinematic viscosity.

The normalised mean velocity $U$ is given by the log-law \cite{bib20,bib5}:

\begin{equation}
\label{eq:mean_vel}
\frac{U_{\rm g,mean}(z)}{u_{\tau}} = \frac{1}{\kappa_g} \log(z^+) + A_{\rm g,mean},
\end{equation}

\noindent where the subscript “g” indicates “global”, $A_{\rm g,mean}=1.01$ \cite{bib5} and the global von K\'arm\'an number $\kappa_g$ is a function of $Re_{\tau}$ \cite{bib8}; $z$ is the distance from the wall and $z^+=z u_{\tau}/\nu_{\rm kin}$ is the normalised distance from the wall. Note that $z/\delta = z^+/Re_{\tau}$.

\begin{figure}[H]
\includegraphics[width=6.5cm]{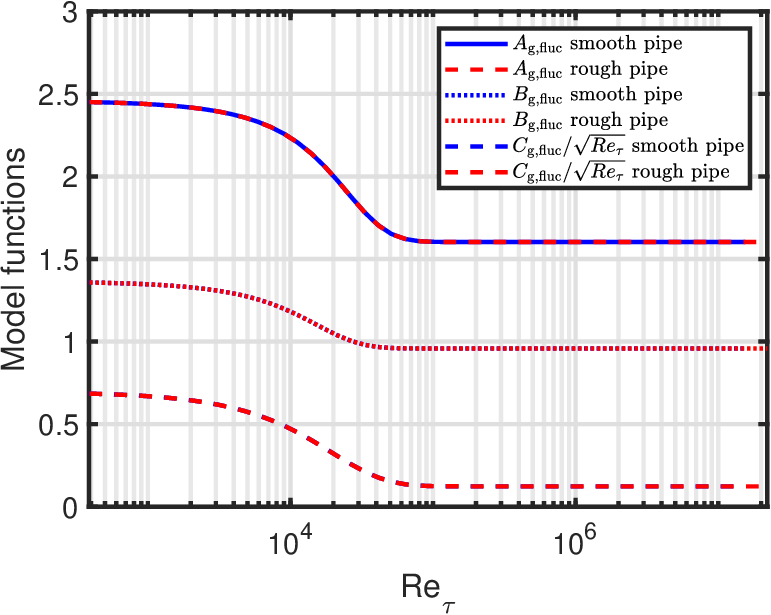}
\hspace{0.3cm}
\includegraphics[width=6.5cm]{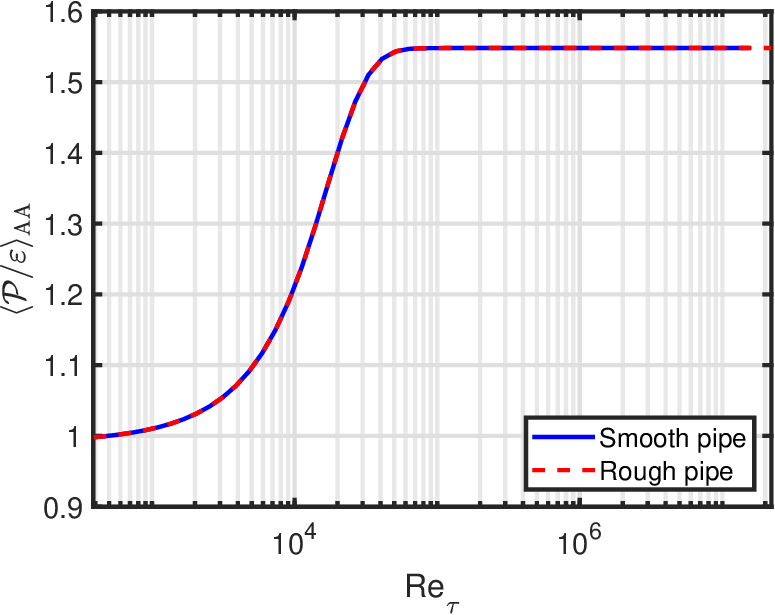}
\caption{Left$-$hand plot: Functions for the square of the normalised fluctuating velocity as a function of friction Reynolds number; right$-$hand plot: area-averaged (AA) turbulence production-to-dissipation ratio as a function of friction Reynolds number. For both plots, the smooth and rough pipe lines are identical and cannot be distinguished.}
\label{fig:fluc_fcts}
\end{figure}

We use an equation for the square of the normalised fluctuating velocity $u$, including the viscous term $V$ as formulated in \cite{bib24}:
\begin{align}
\label{eq:fluc_vel}
\frac{{\overline{u^2_{\rm g,fluc}}}(z)}{u_{\tau}^2} & = B_{\rm g,fluc} - A_{\rm g,fluc} \log (z/\delta) + V(z^+) \\
   & = B_{\rm g,fluc} - A_{\rm g,fluc} \log (z/\delta) - C_{\rm g,fluc} (z^+)^{-1/2}  \\
   & = B_{\rm g,fluc} - A_{\rm g,fluc} \log (z/\delta) -\frac{C_{\rm g,fluc}}{\sqrt{Re_{\tau}}} \sqrt{\frac{\delta}{z}},
\end{align}

\noindent where the subscript “fluc” indicates “fluctuating”. Overbar is time averaging; $A_{\rm g,fluc}$, $B_{\rm g,fluc}$ and $C_{\rm g,fluc}$ are functions of $Re_{\tau}$ \cite{bib6}; see the left$-$hand plot in Figure \ref{fig:fluc_fcts}. Note that we show $C_{\rm g,fluc}/\sqrt{Re_{\tau}}$ instead of $C_{\rm g,fluc}$. These functions are assumed to be identical for smooth and rough pipe flows.

For all figures including results from the model developed in this paper, we include both smooth and rough pipe plots. For many cases, these will be identical, but for several important quantities, they will differ. To make it absolutely clear to the reader when wall roughness impacts the model, we have chosen to include both plots throughout. Details on the smooth and rough pipe settings are provided in Section \ref{sec:mod_results}.

As discussed in \cite{bib6}, we combine our results with findings from \cite{bib12} to derive an area-averaged (AA) turbulence production-to-dissipation ratio:

\begin{equation}
\bigg \langle \frac{\mathcal{P}}{\varepsilon} \bigg \rangle_{\rm AA} = \exp (1.49 - B_{\rm g,fluc}/0.91),
\end{equation}

\noindent with asymptotic limits:

\begin{align}
  \lim_{Re_{\tau}\to0}      \bigg \langle \frac{\mathcal{P}}{\varepsilon} \bigg \rangle_{\rm AA} &= 0.99 \\
  \lim_{Re_{\tau}\to\infty} \bigg \langle \frac{\mathcal{P}}{\varepsilon} \bigg \rangle_{\rm AA} &= 1.55,
\end{align}

\noindent see the right$-$hand plot in Figure \ref{fig:fluc_fcts}. The AA definition can be found in \cite{bib8}.

If turbulence production matches dissipation, the flow is said to be in (local) equilibrium. This assumption is usually made for standard turbulent (eddy) viscosity models. As can be seen in the right$-$hand plot in Figure \ref{fig:fluc_fcts}, the AA production-to-dissipation ratio is close to one for low Reynolds numbers, which can be considered as a flow in (global) equilibrium. However, as the Reynolds number increases, the turbulence production becomes much larger than the turbulence dissipation, which means that the flow will not be in equilibrium. This leads to a need for an investigation on how non-equilibrium flows can be included in turbulence models. A first step in this direction---
an algebraic turbulence model---is the main topic of this paper.

In the remainder of the paper, we will primarily use the expressions for the fluctuating part of the velocity; therefore, we drop the subscript “fluc”; however, the subscript “mean” will be used when we are treating the mean velocity.

\section{Model Overview}
\label{sec:model_overview}

Here, we summarise the main components of our model; for details and derivations, we refer to the SI \cite{bib8}.

\subsection{Basic Model}
\label{subsec:basic}

Simple shear flow is treated, where the mean shear rate (mean velocity gradient) is given by:

\begin{equation}
\mathcal{S}=\partial U/ \partial z,
\end{equation}

\noindent which can be inverted to define a mean shear time scale:

\begin{equation}
\label{eq:s_time}
\tau_{\mathcal{S}} = \frac{1}{\mathcal{S}}
\end{equation}

For equlibrium flows, we use the Prandtl (subscript “P”) characteristic velocity, and for non-equilibrium flows, we use the Kolmogorov--Prandtl (subscript “K-P”) characteristic velocity \cite{bib8}. For the turbulent viscosity and the Reynolds (shear) stress of the streamwise fluctuating velocity $u$ and the wall-normal fluctuating velocity $v$, we write:

\begin{equation}
\nu_{t,{\rm K-P}} = \ell_m^2 | \mathcal{S} | \left( \frac{\mathcal{P}}{\varepsilon} \right)^{-1/2} = \nu_{t,{\rm P}} \left( \frac{\mathcal{P}}{\varepsilon} \right)^{-1/2}
\end{equation}

\begin{equation}
\label{eq:R_stress}
-\overline{uv}_{\rm K-P} = \ell_m^2 \mathcal{S} | \mathcal{S} | \left( \frac{\mathcal{P}}{\varepsilon} \right)^{-1/2} = -\overline{uv}_{\rm P} \left( \frac{\mathcal{P}}{\varepsilon} \right)^{-1/2},
\end{equation}

\noindent where $\ell_m$ is the mixing length, and the turbulent kinetic energy (TKE) production and dissipation rates are given by:

\begin{equation}
\mathcal{P} = \ell_m^2 |\mathcal{S}|^3 \left( \frac{\mathcal{P}}{\varepsilon} \right)^{-1/2}
\end{equation}

\begin{equation}
\varepsilon = \ell_m^2 |\mathcal{S}|^3 \left( \frac{\mathcal{P}}{\varepsilon} \right)^{-3/2}
\end{equation}

The turbulence model constant $C_{\mu}$, which relates the turbulent viscosity, the TKE ($k$) and the dissipation of the TKE, can be written as:

\begin{equation}
\label{eq:C_mu}
C_{\mu} = \frac{\ell_m^4 \mathcal{S}^4}{k^2} \left( \frac{\mathcal{P}}{\varepsilon} \right)^{-2}
\end{equation}

A turbulent length scale can be defined from $k$ and $\varepsilon$ as:

\begin{equation}
L = \frac{k^{3/2}}{\varepsilon},
\end{equation}

\noindent and the corresponding ratio between the mixing length and this new length scale is:

\begin{equation}
\frac{\ell_m}{L} = C_{\mu}^{3/4}
\end{equation}

A turbulent time scale can be defined as:

\begin{equation}
\label{eq:tau_L}
\tau_{L} = \frac{k}{\varepsilon} = \frac{L}{\sqrt{k}}
\end{equation}

The turbulent viscosity ratio is defined as the ratio between the turbulent and kinematic viscosities:

\begin{equation}
\nu_r = \frac{\nu_t}{\nu_{\rm kin}}
\end{equation}

\subsection{Turbulent Mixing Length Scales}
\label{subsec:l_mix_defs}

In \cite{bib8}, we show that the global von K\'arm\'an number $\kappa_g$ transitions from a lower value to a higher value with increasing $Re_{\tau}$. Three mixing length definitions are considered and the decision is made to proceed with the Gersten--Herwig (subscript “G-H”) expression. Taking the AA of the local mixing length, we obtain:

\begin{equation}
\label{eq:l_mix_G-H_AA}
  \langle \ell_{m,{\rm G-H}} \rangle_{\rm AA} = 0.14 \kappa_g \times \delta
\end{equation}

As a shorthand, we can also define the AA mean velocity as:

\begin{equation}
U_m = \langle U_{\rm g,mean} \rangle_{\rm AA}
\end{equation}

\subsection{Turbulence Intensity}
\label{subsec:TI}

We introduce the friction factor $\lambda$ through an equation relating it to the friction velocity and the AA mean velocity:

\begin{equation}
\label{eq:ff}
u_{\tau}^2=\frac{\lambda}{8} \times U_m^2,
\end{equation}

\noindent with $\lambda = 4C_f$, where $C_f$ is the skin friction coefficient. However, we note that Equation (\ref{eq:ff}) is not completely accurate for the measurements used; see \cite{bib6}.

The TKE is equal to the sum of the contributions from streamwise, wall-normal and spanwise velocity fluctuations. We will assume that the TKE is proportional to the square of the streamwise velocity fluctuations:

\begin{equation}
\label{eq:TKE_def}
k=\beta \overline{u^2} = \beta \times \frac{\overline{u^2}}{u_{\tau}^2} \times u_{\tau}^2 = \beta \times \frac{\overline{u^2}}{u_{\tau}^2} \times \frac{\lambda}{8} \times U_m^2,
\end{equation}

\noindent where $\beta$ is a constant of proportionality, see Section \ref{subsec:turb_iso} for more details.

The square of the AA turbulence intensity (TI) is defined as:

\begin{align}
  I_{\rm AA}^2 & = \frac{\langle \overline{u^2} \rangle_{\rm AA}}{U_m^2} = \frac{\langle \overline{u^2} \rangle_{\rm AA}}{u_{\tau}^2} \times \frac{u_{\tau}^2}{U_m^2} = \frac{\langle \overline{u^2} \rangle_{\rm AA}}{u_{\tau}^2} \times \frac{\lambda}{8} \\
   & = \left[ B_g + \frac{3}{2} A_g - \frac{8 C_g}{3 \sqrt{Re_{\tau}}} \right] \times \frac{\lambda}{8},
\end{align}

\noindent with a corresponding AA TKE:

\begin{align}
\label{eq:k_AA_def}
  k_{\rm AA} & = \beta U_m^2 I_{\rm AA}^2 = \beta \langle \overline{u^2} \rangle_{\rm AA} \\
   & = \beta \left[ B_g + \frac{3}{2} A_g - \frac{8 C_g}{3 \sqrt{Re_{\tau}}} \right] \times \frac{\lambda}{8} \times U_m^2
\end{align}

\section{Model Results}
\label{sec:mod_results}

We now apply the results from the preceding sections and the SI to the case of the Princeton Superpipe.

For the TKE, $\beta=1$ and $\beta=1.5$ will be compared; see Section \ref{subsec:turb_iso} for a discussion of these choices.

\subsection{Model Output}
\label{subsec:model_output}

\begin{figure}[H]
\includegraphics[width=6.5cm]{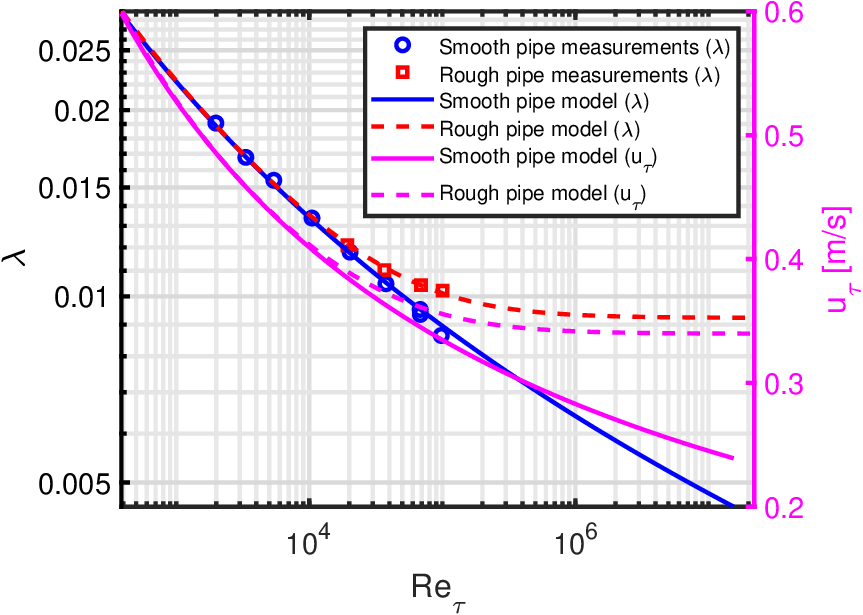}
\hspace{0.3cm}
\includegraphics[width=6.5cm]{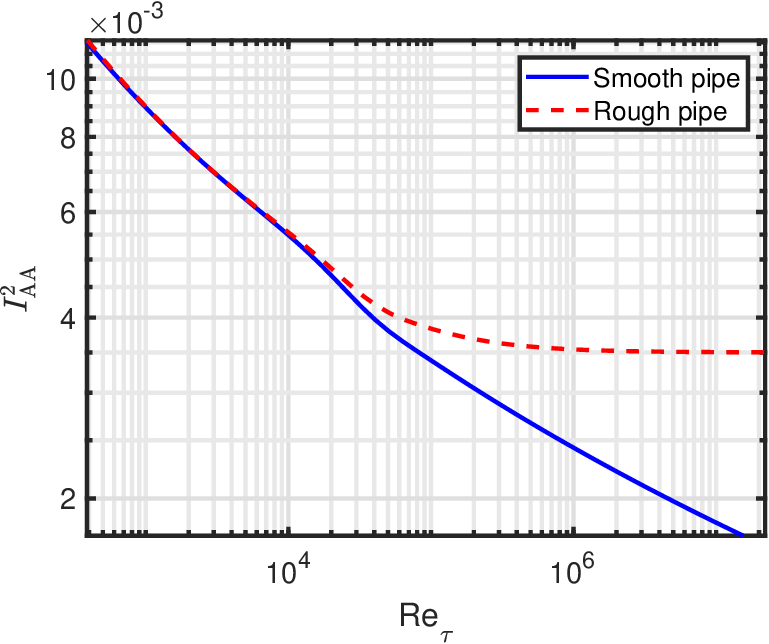}
\caption{Left$-$hand 
 \textls[-15]{plot: Friction factor and friction velocity as a function of $Re_{\tau}$. For the friction factor, Superpipe measurements are included; right$-$hand plot: AA turbulence intensity as a function of $Re_{\tau}$.}}
\label{fig:ff_and_u_tau_vs_Re_tau}
\end{figure}

Outputs from our model will be shown for both smooth and rough pipes. The two initial quantities derived are the friction factor and the friction velocity:

\begin{enumerate}
  \item $\lambda$ from \cite{bib4};
  \item $u_{\tau}$ from Equation (\ref{eq:ff}),
\end{enumerate}

\noindent where the friction velocity enables the calculation of the friction Reynolds number:

\begin{equation}
Re_{\tau} = \frac{R u_{\tau}}{\nu_{\rm kin}},
\end{equation}

\noindent see the left$-$hand plot in Figure \ref{fig:ff_and_u_tau_vs_Re_tau}. It is clear that the smooth and rough pipe results begin to deviate above $Re_{\tau} \sim 10^4$.
\vspace{-6pt}

Having calculated the friction factor, we can derive the AA TI as:

\begin{equation}
\label{eq:I_AA_output}
I_{\rm AA}^2 = \left[ B_g + \frac{3}{2} A_g - \frac{8 C_g}{3 \sqrt{Re_{\tau}}} \right] \times \frac{\lambda}{8},
\end{equation}

\noindent see the right$-$hand plot in Figure \ref{fig:ff_and_u_tau_vs_Re_tau}. Again, the smooth and rough pipe friction factor difference is reflected in the divergence of the TIs: The smooth-pipe TI continues to decrease, whereas the rough-pipe TI reaches a plateau.

\subsubsection{Quantities Depending on TKE, but Not Mixing Length}
\label{subsubsec:tke_no_length}

Now, we split our analysis in two parts: in this section, expressions depend on $\beta$ (TKE) but not on $\langle \ell_{m,{\rm G-H}} \rangle_{\rm AA}$. In Section \ref{subsubsec:length_no_tke}, we study the opposite case.

\begin{figure}[H]
\includegraphics[width=7.5cm]{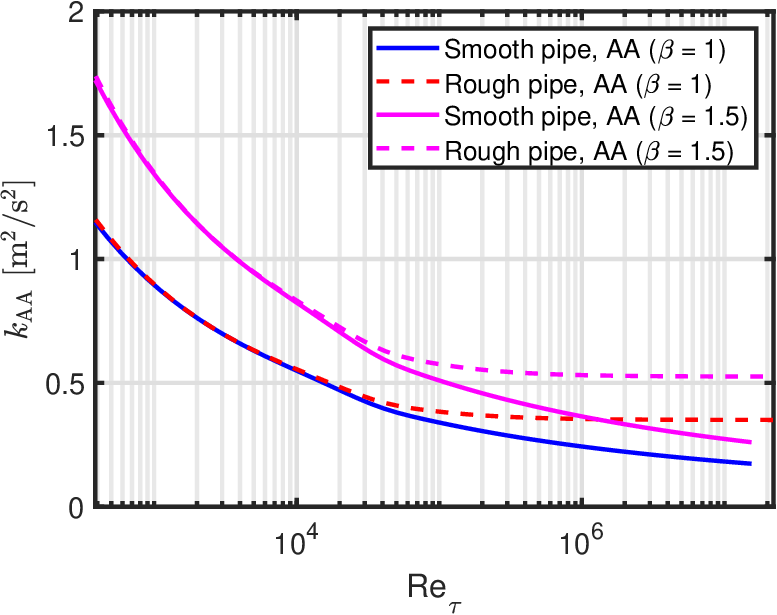}
\caption{Turbulent kinetic energy as a function of $Re_{\tau}$ for $\beta=1$ and $\beta=1.5$.}
\label{fig:k_AA_vs_Re_tau}
\end{figure}

The TKE is defined as:

\begin{equation}
k_{\rm AA} = \beta \left[ B_g + \frac{3}{2} A_g - \frac{8 C_g}{3 \sqrt{Re_{\tau}}} \right] \times \frac{\lambda}{8} \times U_m^2,
\end{equation}

\noindent see Figure \ref{fig:k_AA_vs_Re_tau}. The two different $\beta$ values lead to an overall shift of the TKE. The rough-pipe TKE reaches a fixed value for high Reynolds numbers in contrast to the smooth-pipe TKE, which continues to decrease.

\begin{figure}[H]
\includegraphics[width=6cm]{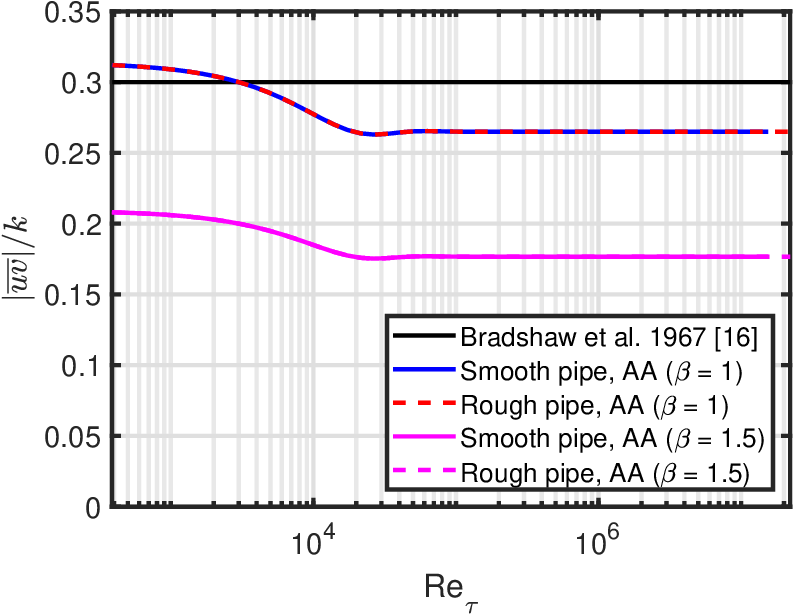}
\hspace{0.3cm}
\includegraphics[width=6cm]{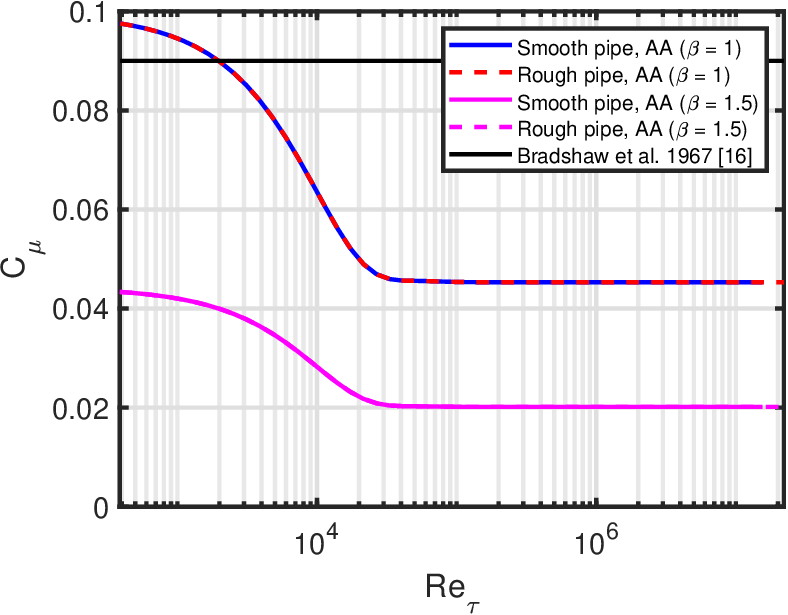}
\caption{Left$-$hand plot: The ratio of the absolute value of the Reynolds stress to the turbulent kinetic energy as a function of $Re_{\tau}$ for $\beta=1$ and $\beta=1.5$; right$-$hand plot: $C_{\mu}$ as a function of $Re_{\tau}$ for $\beta=1$ and $\beta=1.5$. For both plots, smooth and rough pipe lines are identical and cannot be \mbox{distinguished.}
}
\label{fig:Bradshaw_cst_vs_Re_tau}
\end{figure}

The ratio of the absolute value of the Reynolds stress to the TKE is:

\begin{equation}
\frac{|\overline{uv}|_{\rm AA}}{k_{\rm AA}} = \frac{u_{\tau}^2}{k_{\rm AA}} \times \bigg \langle \frac{\mathcal{P}}{\varepsilon} \bigg \rangle_{\rm AA}^{-1/2} = \frac{1}{\beta \times \left( B_g + \frac{3}{2} A_g - \frac{8 C_g}{3 \sqrt{Re_{\tau}}} \right)} \times \bigg \langle \frac{\mathcal{P}}{\varepsilon} \bigg \rangle_{\rm AA}^{-1/2},
\end{equation}

\noindent which is compared to the standard value of 0.3 \cite{bib9,bib8} in the left$-$hand plot in Figure \ref{fig:Bradshaw_cst_vs_Re_tau}. The magnitude of our AA expressions match the standard value quite well for the $\beta=1$ case, but it does decrease by 15\% across the high Reynolds number transition region, mainly due to the scaling with the turbulence production-to-dissipation ratio. In contrast, the magnitude of our AA expressions with $\beta=1.5$ is much smaller than the standard value. See Section \ref{subsec:scal_C_mu} for a discussion on these findings.

Note also that there is no difference between the smooth and rough pipe models.

The AA turbulence model constant $C_{\mu}$ is defined as:

\begin{align}
  C_{\mu,{\rm AA}} & = \frac{u_{\tau}^4}{k_{\rm AA}^2}  \times \bigg \langle \frac{\mathcal{P}}{\varepsilon} \bigg \rangle_{\rm AA} ^{-2} \\ & = \frac{1}{\beta^2 \times \left( \frac{\langle \overline{u^2} \rangle_{\rm AA}}{u_{\tau}^2} \right)^2} \times \bigg \langle \frac{\mathcal{P}}{\varepsilon} \bigg \rangle_{\rm AA} ^{-2} \\
   & = \frac{1}{\beta^2 \times \left( B_g + \frac{3}{2} A_g - \frac{8 C_g}{3 \sqrt{Re_{\tau}}} \right)^2  \times \Big \langle \frac{\mathcal{P}}{\varepsilon} \Big \rangle_{\rm AA}^{2}},
   \label{eq:C_mu_AA}
\end{align}

\noindent see the right$-$hand plot in Figure \ref{fig:Bradshaw_cst_vs_Re_tau}. For $\beta=1$, the values at low Reynolds numbers are fairly close to the standard value of 0.09, but they decrease to around half of that value for high Reynolds numbers. An increase in $\beta$ to 1.5 leads to a downward shift of $C_{\mu}$. As above, we refer to Section \ref{subsec:turb_iso} for a discussion of these findings.

\begin{figure}[H]
\includegraphics[width=6.5cm]{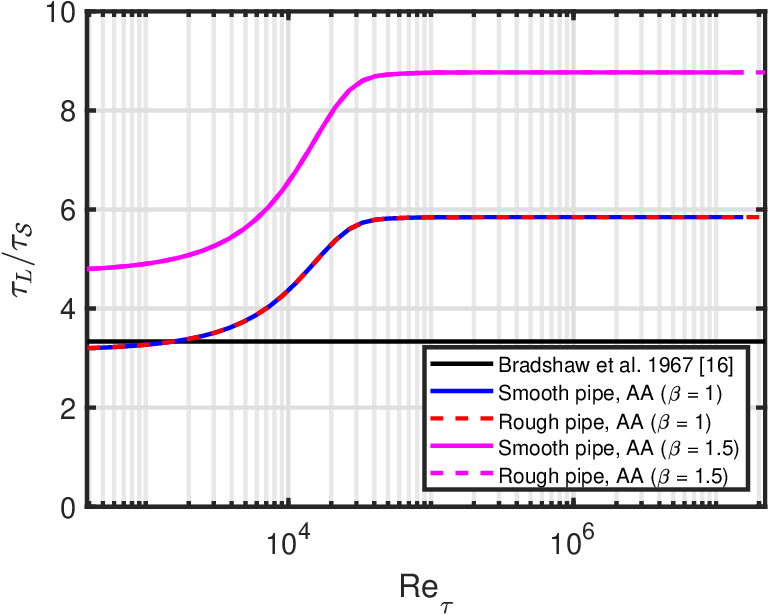}
\hspace{0.3cm}
\includegraphics[width=6.5cm]{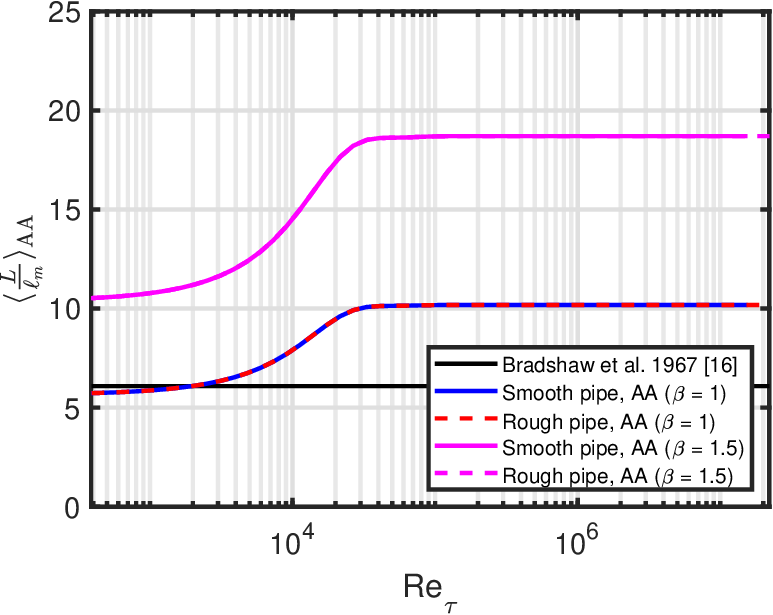}
\caption{Left$-$hand plot: The turbulence-to-mean shear time scale ratio as a function of $Re_{\tau}$ for $\beta=1$ and $\beta=1.5$. Right$-$hand plot: AA length scale ratio as a function of $Re_{\tau}$ for $\beta=1$ and $\beta=1.5$. For both plots, smooth and rough pipe lines are identical and cannot be distinguished. 
}
\label{fig:tau_L_div_tau_S}
\end{figure}

The AA definition of the turbulence-to-mean shear time scale ratio is:

\begin{equation}
\bigg \langle \frac{\tau_L}{\tau_{\mathcal{S}}} \bigg \rangle_{\rm AA} = \frac{\mathcal{S}_{\rm AA} k_{\rm AA}}{\varepsilon_{\rm AA}} = \frac{\langle L \rangle_{\rm AA}}{\sqrt{k_{\rm AA}}} \frac{u_{\tau}}{\langle \ell_m \rangle_{\rm AA}} = \bigg \langle \frac{L}{\ell_m} \bigg \rangle_{\rm AA} \frac{u_{\tau}}{\sqrt{\beta \langle \overline{u^2} \rangle_{\rm AA}}},
\end{equation}

\noindent where we have used Equations (\ref{eq:s_time}) and (\ref{eq:tau_L}); see the left$-$hand plot in Figure \ref{fig:tau_L_div_tau_S}. The standard value of $10/3$ \cite{bib8} is included for reference. For $\beta=1$ and low Reynolds number, the AA definition matches the standard value quite well, but increases for higher Reynolds numbers. For $\beta=1.5$, the AA definition is shifted upwards. Other time scales are discussed in Section \ref{subsec:time_scales}.

Finally, we show the length scale ratio:

\begin{equation}
\bigg \langle \frac{L}{\ell_m} \bigg \rangle_{\rm AA} = C_{\mu,{\rm AA}}^{-3/4} = \beta^{3/2} \times \left( B_g + \frac{3}{2} A_g - \frac{8 C_g}{3 \sqrt{Re_{\tau}}} \right)^{3/2}  \times \bigg \langle \frac{\mathcal{P}}{\varepsilon} \bigg \rangle_{\rm AA}^{3/2},
\end{equation}

\noindent see the right$-$hand plot in Figure \ref{fig:tau_L_div_tau_S}. The standard value of 6.1 $\left( 0.09^{-3/4} \right)$ is included for reference. Overall, we conclude that $\langle L \rangle_{\rm AA} \sim 6-19 \times \langle \ell_m \rangle_{\rm AA}$, depending on the Reynolds number and $\beta$ values. Specifically applied to the Gersten--Herwig mixing length, we have:

\begin{equation}
\langle L_{\rm G-H} \rangle_{\rm AA} \sim 6-19 \times \langle \ell_{m, {\rm G-H}} \rangle_{\rm AA} \sim 0.3-1 \times \delta,
\end{equation}

\noindent leading to an interpretation of $L$ as being a characteristic length corresponding to the boundary layer thickness.

\subsubsection{Quantities Depending on Mixing Length, but Not TKE}
\label{subsubsec:length_no_tke}

We now treat the inverse case of what was covered in Section \ref{subsubsec:tke_no_length}, where the quantities depend on $\langle \ell_{m,{\rm G-H}} \rangle_{\rm AA}$ but not on $\beta$ (TKE).

We start by writing expressions for the TKE production and dissipation rates:

\begin{equation}
\mathcal{P}_{\rm AA} = \frac{u_{\tau}^3}{\langle \ell_{m,{\rm G-H}} \rangle_{\rm AA}} \times \bigg \langle \frac{\mathcal{P}}{\varepsilon} \bigg \rangle_{\rm AA} ^{-1/2}
\end{equation}

\begin{equation}
\varepsilon_{\rm AA} = \frac{u_{\tau}^3}{\langle \ell_{m,{\rm G-H}} \rangle_{\rm AA}} \times \bigg \langle \frac{\mathcal{P}}{\varepsilon} \bigg \rangle_{\rm AA}^{-3/2}
\end{equation}

\begin{figure}[H]
\includegraphics[width=8cm]{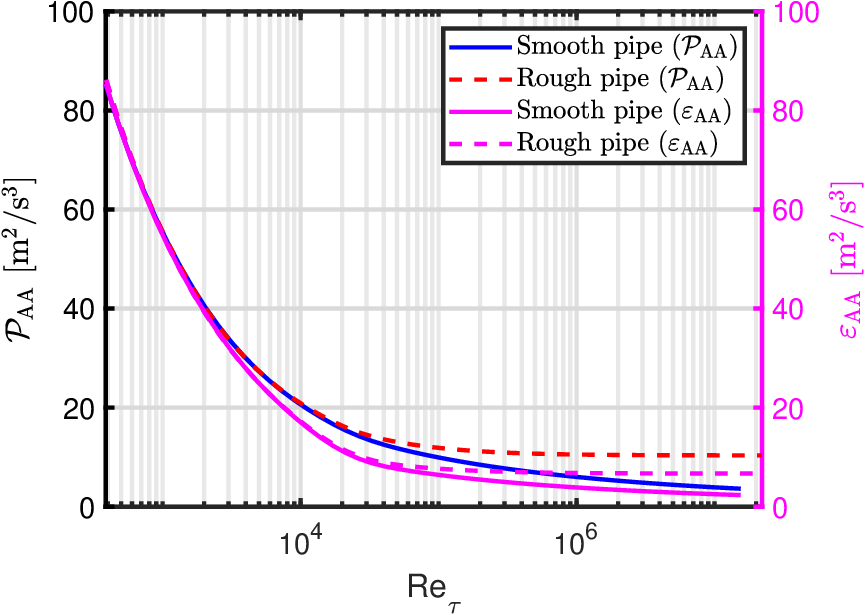}
\caption{Turbulent kinetic energy production and dissipation rate as a function of $Re_{\tau}$.}
\label{fig:production_and_dissipation_vs_Re_tau}
\end{figure}

\textls[-20]{The TKE production and dissipation rates are shown in Figure \ref{fig:production_and_dissipation_vs_Re_tau}. The dominating quantity is $u_{\tau}^3$, which leads to a rapid decrease with increasing Reynolds number. Further, the smooth pipe values continue to decrease while the rough pipe values reach a constant number.}

\begin{figure}[H]
\includegraphics[width=6.5cm]{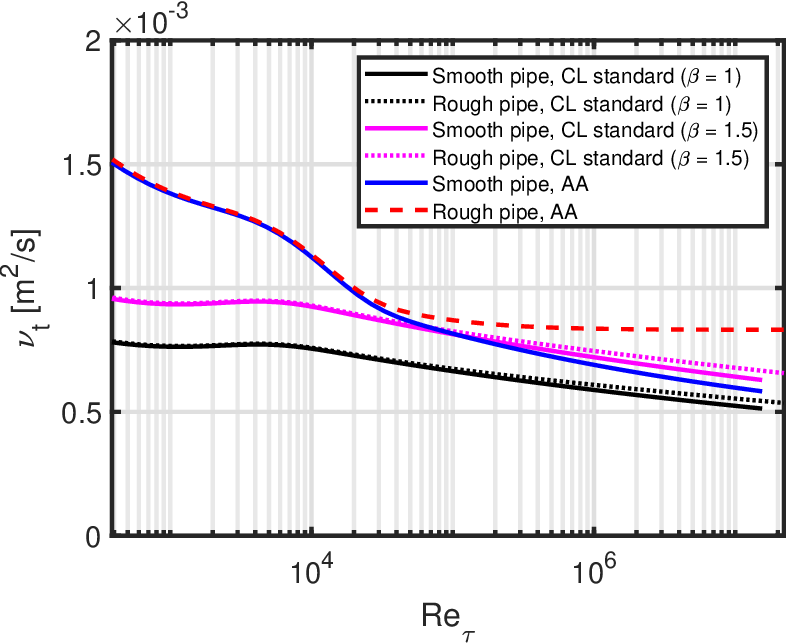}
\hspace{0.3cm}
\includegraphics[width=6.5cm]{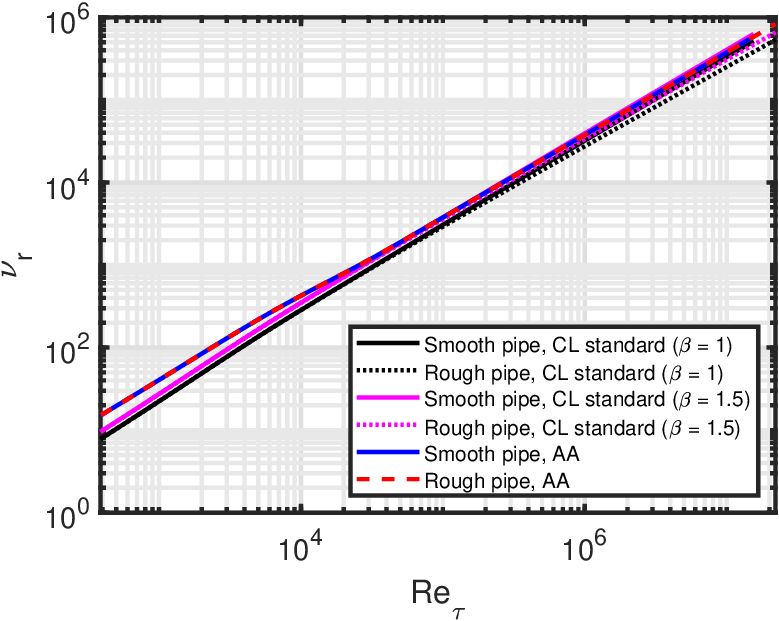}
\caption{Turbulent 
 viscosity (left$-$hand plot) and turbulent viscosity ratio (right$-$hand plot) as a function of $Re_{\tau}$ for $\beta=1$ and $\beta=1.5$.}
\label{fig:nu_t_vs_Re_tau}
\end{figure}

The turbulent viscosity is given by:

\begin{equation}
\nu_{t,{\rm AA}} = u_{\tau} \langle \ell_{m,{\rm G-H}} \rangle_{\rm AA} \times \bigg \langle \frac{\mathcal{P}}{\varepsilon} \bigg \rangle_{\rm AA}^{-1/2} = C_{\mu,{\rm AA}} \times \frac{k_{\rm AA}^2}{\varepsilon_{\rm AA}},
\end{equation}

\noindent see the left$-$hand plot in Figure \ref{fig:nu_t_vs_Re_tau}, with corresponding turbulent viscosity ratios in the right$-$hand plot in Figure \ref{fig:nu_t_vs_Re_tau}. The values from our model do not depend on $\beta$, but for the lines marked “CL standard”, there is a dependency, which will be discussed in Section \ref{subsec:scal_nu_t}.

For our model at low Reynolds numbers, the turbulent viscosity decreases for increasing Reynolds numbers for both smooth and rough pipes. For high Reynolds numbers, the smooth-pipe turbulent viscosity continues to decrease, while the rough-pipe turbulent viscosity approaches a constant value. For both smooth and rough pipes, there is a transition region which is due to the combination of the Reynolds number dependency of the von K\'arm\'an number and the turbulence production-to-dissipation ratio.

The turbulent viscosity ratio covers a large range from about 10 to $10^6$, which is directly linked to the range of the kinematic viscosities \cite{bib8}.

\section{Discussion}
\label{sec:disc}

\subsection{Turbulence Isotropy}
\label{subsec:turb_iso}

$\beta$ as defined in Equation (\ref{eq:TKE_def}) is 1.5 for isotropic turbulence, where each of the three fluctuating velocity components contribute equally. For actual flows, what is typically observed is that half of the TKE is in the streamwise fluctuations and the other half is in the sum of the wall-normal and spanwise fluctuations, which implies a $\beta$ of 1 \cite{bib13,bib34}.

An open question is whether $\beta$ is a function of $Re_{\tau}$ or if variations in $\mathcal{P}/\varepsilon$ (based on streamwise fluctuations) are related to a change in $\beta$?

We will use a Reynolds number independent $\beta=1$ for all figures in Section \ref{sec:disc}, but note that this is an assumption which can,
and should, be questioned.

\subsection{Physical Mechanism}

One explanation for the high Reynolds number transition region is an increase in large-scale structures in the wake region. This can be thought of as an analogy to the “drag \mbox{crisis” \cite{bib28}.}

The question is if these structures are active or inactive, i.e., if they contribute to the turbulent shear stress or not.

In \cite{bib9}, the ratio of the absolute value of the Reynolds stress to the TKE is discussed: “In the last-named paper, evidence is presented that the considerable variation in $a_1$, observed experimentally is at least partly due to an ‘inactive’, quasi-irrotational component of the turbulent motion (Townsend 1961), which does not contribute to the shear stress or the dissipation and can therefore be disregarded; therefore, $a_1$ = constant is a much better approximation than at first appears.” In our notation, $2a_1=|\overline{uv}|/k$ and it is thus argued that this quantity can be considered constant.

A similar interpretation is provided in \cite{bib11}, which cites results from \cite{bib25}: “[...]which argued that the excess of turbulent production in the log layer feeds inactive motions that do not contribute to the turbulent shear stress, but transfer energy to other locations of \mbox{the flow”.}

An opposing view can be found in \cite{bib10}, where it is stated that very-large-scale motions (VLSM) “[...]can contain up to 60\% of the cumulative fraction of the Reynolds shear stress[...]”.

To summarise, it remains unclear to which extent the high-Reynolds-number structures contribute to the turbulent shear stress. A possible reconciliation of these views is presented in \cite{deshpande2023}, where VLSM (or superstructures) are interpreted to be inactive structures formed as a concatenation of active eddies.

An additional uncertainty of our analysis is that it is based on the Princeton Superpipe measurements, where it has been found that the inner peak of the streamwise fluctuations is not resolved for all Reynolds numbers \cite{bib35}. However, our main results should be robust, since the global (integral) treatment is not dominated by this inner peak.

\subsection{Scaling of \texorpdfstring{$C_{\mu}$}{-}}
\label{subsec:scal_C_mu}

Scaling of $C_{\mu}$ with $\mathcal{P}/\varepsilon$ has been discussed in \cite{bib31}. We note that the definition of $\mathcal{P}/\varepsilon$ in \cite{bib31} is different from our ours; in \cite{bib31}, $\mathcal{P}/\varepsilon$ is weighted with $|\overline{uv}|$, whereas we use area-averaging. An expression for $C_{\mu}$ as a function of $\mathcal{P}/\varepsilon$, valid for $\mathcal{P}/\varepsilon > 1$, is given as:

\begin{equation}
C_{\mu,{\rm R}} = \frac{2}{3} \times \frac{1-\alpha_0}{\omega_0} \times \frac{1-\frac{1}{\omega_0}(1-\alpha_0 \langle \mathcal{P}/\varepsilon \rangle_{\rm AA})}{\left[ 1 + \frac{1}{\omega_0} (\langle \mathcal{P}/\varepsilon \rangle_{\rm AA}-1) \right]^2},
\end{equation}

\noindent with:

\begin{align}
  \alpha_0 & = 2.8 \\
  \omega_0 & = 0.55,
\end{align}

\noindent where we use the subscript “R” for Rodi and have replaced the turbulence production-to-dissipation ratio from \cite{bib31} with our area-averaged definition.

As written in \cite{bib31}, under certain conditions “[...]the mixing length hypothesis implies that $\mathcal{P}/\varepsilon$ is constant; but $\mathcal{P}/\varepsilon$ need not equal unity”.

\begin{figure}[H]
\includegraphics[width=6.5cm]{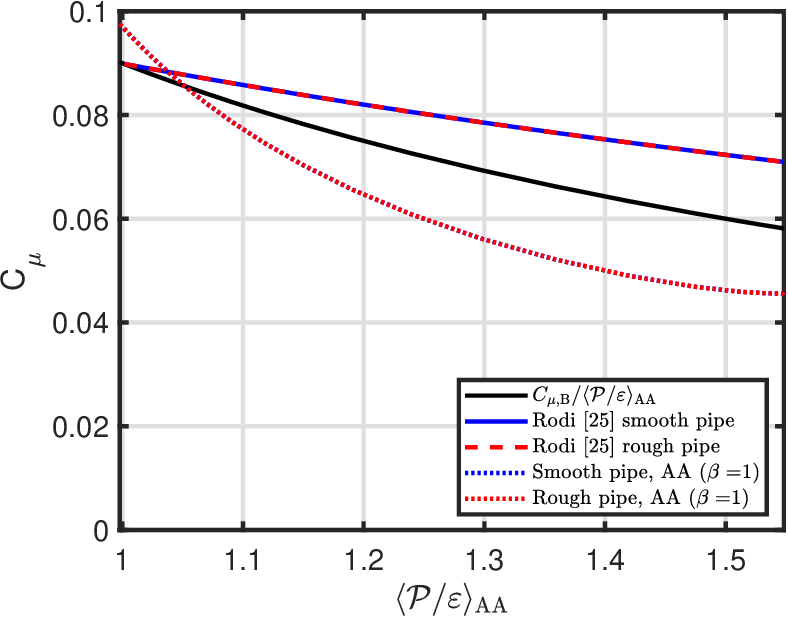}
\hspace{0.3cm}
\includegraphics[width=6.5cm]{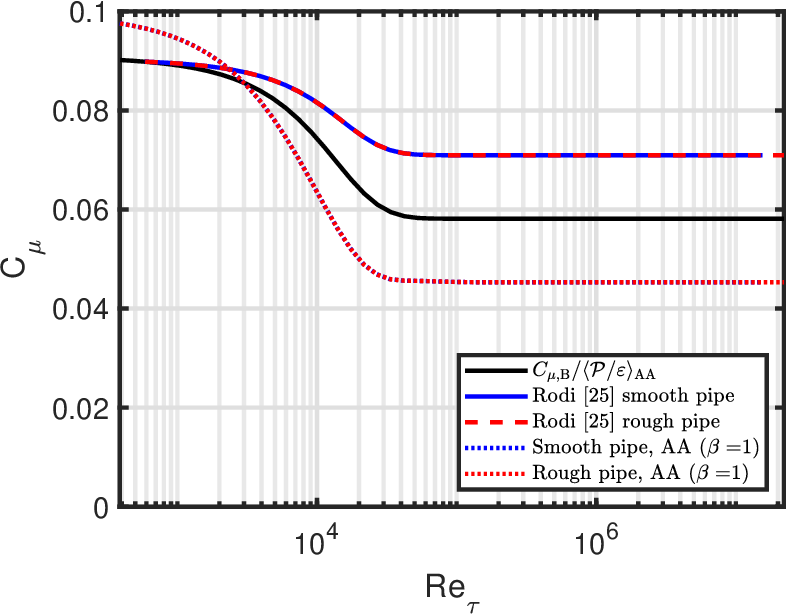}
\caption{$C_{\mu,{\rm R}}$ from \protect\cite{bib31} compared to $C_{\mu,{\rm B}}/\langle \mathcal{P}/\varepsilon \rangle_{\rm AA}$ and our AA expressions. Left$-$hand plot: As a function of $\langle \mathcal{P}/\varepsilon \rangle_{\rm AA}$. Right$-$hand plot: As a function of $Re_{\tau}$. Smooth and rough pipe lines are identical and cannot be distinguished.}
\label{fig:Rodi_C_mu}
\end{figure}

In Figure \ref{fig:Rodi_C_mu}, $C_{\mu,{\rm R}}$ from \cite{bib31} is compared to $C_{\mu,{\rm B}}/\langle \mathcal{P}/\varepsilon \rangle_{\rm AA}$ and our AA expressions.

The difference in scaling with $\mathcal{P}/\varepsilon$ between previous findings and our results originates from the assumption of whether $|\overline{uv}|$ scales with $\mathcal{P}/\varepsilon$ or not. Overall, previous findings predict a scaling of $C_{\mu}$ with $(\mathcal{P}/\varepsilon)^{-1}$, while our expressions imply a scaling of $C_{\mu}$ with $(\mathcal{P}/\varepsilon)^{-2}$. In \cite{bib31}, the weighting of $\mathcal{P}/\varepsilon$ with $|\overline{uv}|$ somewhat complicates the comparison.

\subsection{Scaling of \texorpdfstring{$\nu_t$}{-} and \texorpdfstring{$I$}{-}}
\label{subsec:scal_nu_t}

The scaling of the turbulent viscosity can be compared to the standard CL expression from \cite{bib14}, which we have modified to include $\beta$ specifically:

\begin{equation}
\label{eq:nu_t_CL_std}
\nu_{t,{\rm CL}} = \sqrt{\beta} C_{\mu,{\rm B}}^{1/4} \times U_{\rm CL} I_{\rm CL} \ell_{m,{\rm G-H,CL}},
\end{equation}

\noindent where we use \cite{bib33}:

\begin{equation}
I_{\rm CL} = 0.055 \times Re_D^{-0.041}
\end{equation}

In Figure \ref{fig:nu_t_vs_Re_tau} (both plots), we have included the turbulent viscosity from Equation (\ref{eq:nu_t_CL_std}) and marked it “CL standard”. This does not match our model for the low Reynolds number range, but it does match the smooth pipe model expression quite well for the high Reynolds number range and $\beta=1.5$. There is no $\beta$ dependency of the turbulent viscosity in \mbox{our model.}

\subsection{Time Scales}
\label{subsec:time_scales}

From the log-law mean velocity, a single length scale is associated with the von K\'arm\'an mixing length, or rather, a continuum of length scales increases from the wall.

In contrast, the mixing length expressions by Nikuradse and Gersten--Herwig can be considered as consisting of two different ranges, one close to the wall and another one towards the CL. These length scales can be used to define two time scales. A similar conclusion can be drawn from the fluctuations as defined in Equation (\ref{eq:fluc_vel}), which also corresponds to two length scales as noted in \cite{bib6}.

Below, we will explore the idea of two time scales based on the assumption of two corresponding length scales. Note that a different time scale ratio for the turbulence-to-mean shear is contained in our model and has been treated in \cite{bib8}; see \mbox{also Figure \ref{fig:tau_L_div_tau_S}.}

\subsubsection{\texorpdfstring{$k-\varepsilon$}{-} Turbulence Model with Two Time Scales}
\label{subsubsec:turb_two_time}

Turbulence models with multiple time scales have been treated in \cite{bib16}, with more recent further development to be found in \cite{bib21}. We base our discussion on homogeneous flow, where equations for the time evolution of $k$ and $\varepsilon$ can be written as \cite{bib15}:
\begin{align}
\label{eq:k_eq}
  \frac{{\rm d}k}{{\rm d}t} & = \mathcal{P}-\frac{\varepsilon}{c_s} \\
\label{eq:epsilon_eq}
  \frac{{\rm d}\varepsilon}{{\rm d}t} & = \left( C_{\varepsilon 1} \mathcal{P} - C_{\varepsilon 2} \frac{\varepsilon}{c_s} \right) \frac{\varepsilon}{k},
\end{align}

\noindent where the standard values of the coefficients $C_{\varepsilon 1}=1.44$, $C_{\varepsilon 2}=1.92$ \cite{bib22} and:

\begin{equation}
c_s = \frac{\tau_{\rm fluc}}{\tau_{\rm mean}}
\end{equation}

\noindent is the ratio of the turbulence and the mean flow time scales. Note that $c_s = 1$ for the standard $k-\varepsilon$ model \cite{bib19}.

The evolution of the turbulence time scale $\tau_{\rm fluc} = \tau_L = k/\varepsilon$ (see Equation (\ref{eq:tau_L})) is given as \cite{bib15}:

\begin{equation}
\frac{{\rm d} (k/\varepsilon)}{{\rm d} t} = (1-C_{\varepsilon 1}) \frac{\mathcal{P}}{\varepsilon} - \frac{1}{c_s} (1-C_{\varepsilon 2})
\end{equation}

\begin{figure}[H]
\includegraphics[width=6.5cm]{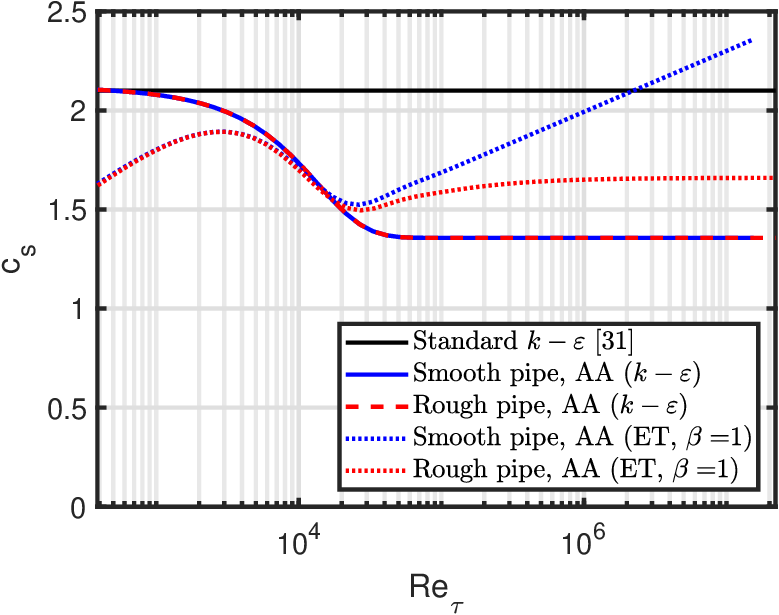}
\hspace{0.3cm}
\includegraphics[width=6.5cm]{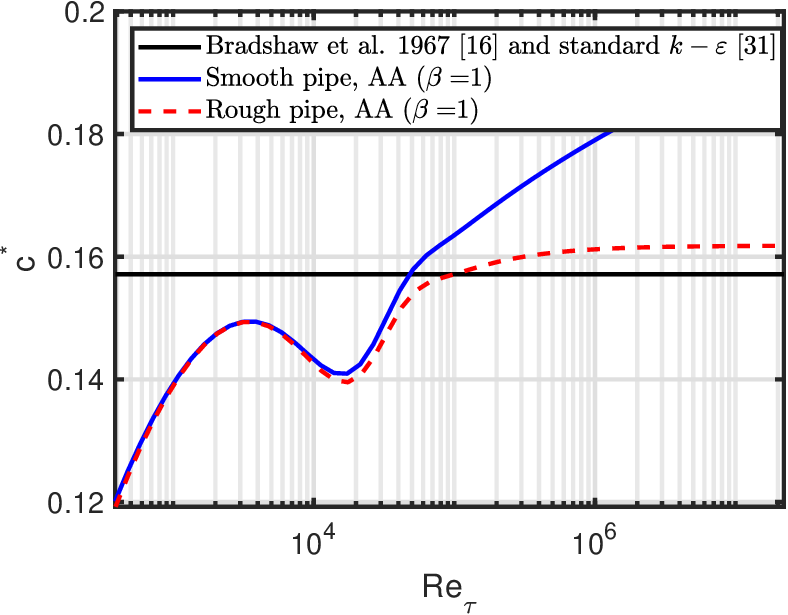}
\caption{Left$-$hand plot: The ratio of the turbulence and the mean flow time scales as a function of $Re_{\tau}$. The smooth and rough pipe lines (for $k-\varepsilon$) are identical and cannot be distinguished. Right$-$hand plot: Exponential constant as a function of $Re_{\tau}$. 
}
\label{fig:c_s_comp}
\end{figure}

If $k/\varepsilon$ does not vary with time, d$(k/\varepsilon)/{\rm d}$t = 0 and we have:

\begin{equation}
c_s = \left( \frac{C_{\varepsilon 2}-1}{C_{\varepsilon 1}-1} \right) \left( \frac{\mathcal{P}}{\varepsilon} \right)^{-1},
\end{equation}

\noindent which can be written for AA as:

\begin{equation}
\langle c_s \rangle_{\rm AA} = \left( \frac{C_{\varepsilon 2}-1}{C_{\varepsilon 1}-1} \right) \bigg \langle \frac{\mathcal{P}}{\varepsilon} \bigg \rangle_{\rm AA}^{-1} = 2.1 \times \bigg \langle \frac{\mathcal{P}}{\varepsilon} \bigg \rangle_{\rm AA}^{-1},
\end{equation}

\noindent where the final equation assumes that $C_{\varepsilon 1}$ and $C_{\varepsilon 2}$ are constant, i.e., do not scale with $Re_{\tau}$ (or, equivalently, with $\mathcal{P}/\varepsilon$). See the left$-$hand plot in Figure \ref{fig:c_s_comp}, where this $k-\varepsilon$ expression is equal to the 2.1 “Standard $k-\varepsilon$” value for low Reynolds numbers and decreases monotonically to 1.4 for high Reynolds numbers.

\subsubsection{Eddy-Turnover Time Scales}

An alternative time scale ratio can be defined by an eddy-turnover (ET) time, both for the fluctuations and for the mean flow \cite{bib27}:
\begin{align}
  \tau_{\rm ET, fluc} & = \frac{\ell_m}{\sqrt{\overline{u^2}}} \\
  \tau_{\rm ET, mean} & = \frac{L}{U},
\end{align}

\noindent along with their ratio:

\begin{equation}
c_{s,{\rm ET}} = \frac{\tau_{\rm ET, fluc}}{\tau_{\rm ET, mean}} = \frac{\ell_m}{L} \frac{U}{\sqrt{\overline{u^2}}}
\end{equation}

For AA, this can be rewritten to:

\begin{equation}
\langle c_{s,{\rm ET}} \rangle_{\rm AA} = \bigg \langle \frac{\ell_m}{L} \bigg \rangle_{\rm AA} \times \frac{1}{\sqrt{I_{\rm AA}^2}},
\end{equation}

\noindent see the left$-$hand plot in Figure \ref{fig:c_s_comp} marked “ET”. The range of $c_s$ values is similar to the ones from Section \ref{subsubsec:turb_two_time}, but not monotonic. The main difference is the increase in $\langle c_{s,{\rm ET}} \rangle_{\rm AA}$ for the smooth pipe, which is due to an increase in the fluctuating time scale, which in turn is due to a decreasing TKE.

\subsubsection{Time Evolution of TKE}

The time scale ratio from Section \ref{subsubsec:turb_two_time} can be used to define an equation for the time evolution of the TKE:

\begin{equation}
\frac{\tau_L}{k} \frac{{\rm d}k}{{\rm d}t} = \frac{\mathcal{P}}{\varepsilon}-\frac{1}{c_s},
\end{equation}

\noindent which has the solution:

\begin{equation}
\label{eq:tke_fct_time}
k(t) = k(0) \exp \left[ \frac{t}{\tau_L} \left( \frac{\mathcal{P}}{\varepsilon}-\frac{1}{c_s} \right) \right],
\end{equation}

\noindent where $k$ grows if:

\begin{equation}
\frac{\mathcal{P}}{\varepsilon} > \frac{1}{c_s}
\end{equation}

By defining normalised quantities $k^*(t)=k(t)/k(0)$ and $t^*=\mathcal{S} t$ \cite{bib37}, we can rewrite Equation (\ref{eq:tke_fct_time}) to:
\begin{align}
  k^*(t^*) & = \exp \left[ t^* \times \frac{1}{\tau_L \mathcal{S}} \left( \frac{\mathcal{P}}{\varepsilon}-\frac{1}{c_s} \right) \right] \\
   & = \exp \left[ t^* \times \frac{k C_{\mu}}{|\overline{uv}|} \left( \frac{\mathcal{P}}{\varepsilon}-\frac{1}{c_s} \right) \right],
\end{align}

\noindent where we have used Equation (\ref{eq:tau_L}) for the second equation.

\begin{figure}[H]
\includegraphics[width=8cm]{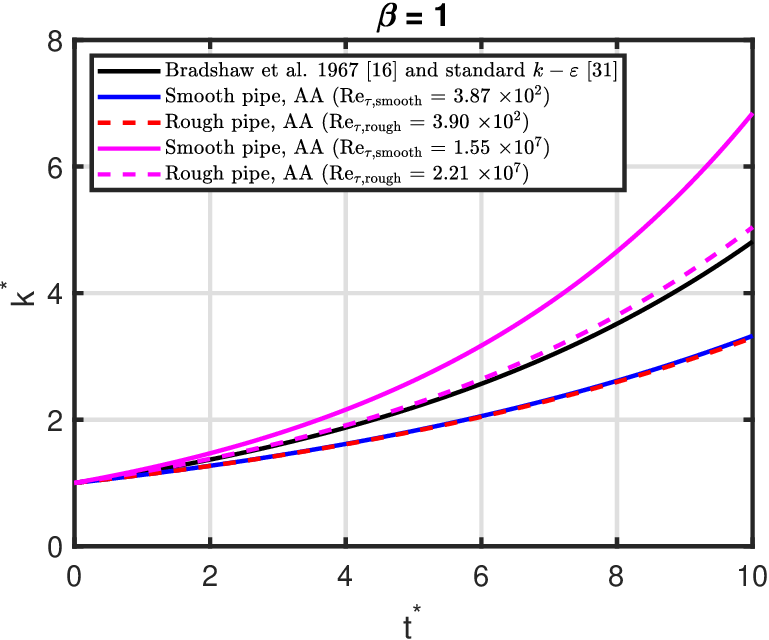}
\caption{Normalised 
 turbulent kinetic energy as a function of normalised time for low and high friction Reynolds numbers. 
}
\label{fig:k_star_vs_t_star}
\end{figure}

For $c_s = 2.1$ and the definitions in \cite{bib8}, $k C_{\mu}/|\overline{uv}|=0.3$, and we can add values to the equation:
\begin{align}
  k_{\rm B}^*(t^*) & = \exp \left[ 0.16 \times t^* \right] \\
   & = \exp \left[ c_{\rm B}^* \times t^* \right],
\end{align}

\noindent where $c_{\rm B}^*$ is shown in the right$-$hand plot in Figure \ref{fig:c_s_comp} and $k_{\rm B}^*$ is shown in Figure \ref{fig:k_star_vs_t_star}.

As for the standard definitions, we can write the TKE as a function of time for our AA model with $c_{s,{\rm ET}}$:
\begin{align}
  k_{\rm AA}^*(t^*) & = \exp \left[ \frac{k_{\rm AA} C_{\mu,{\rm AA}}}{|\overline{uv}|_{\rm AA}} \left( \bigg \langle \frac{\mathcal{P}}{\varepsilon} \bigg \rangle_{\rm AA}-\frac{1}{\langle c_{s,{\rm ET}} \rangle_{\rm AA}} \right) \times t^* \right] \\
   & = \exp \left[ c_{\rm AA}^* \times t^* \right],
\end{align}

\noindent also shown in the right$-$hand plot in Figure \ref{fig:c_s_comp} ($c_{\rm AA}^*$) and in Figure \ref{fig:k_star_vs_t_star} ($k_{\rm AA}^*$).

In the right$-$hand plot in Figure \ref{fig:c_s_comp}, we see that $c_{\rm AA}^*$ is a function of Reynolds numbers as opposed to $c_{\rm B}^*$, which is independent of Reynolds numbers. The increase in $c_{\rm AA}^*$ for the smooth pipe is because of an increase in $\langle c_{s,{\rm ET}} \rangle_{\rm AA}$. It is interesting to note that for high Reynolds numbers, $k_{\rm AA}^*$ increases faster for the smooth pipe than for the rough pipe.

\vspace{-6pt}

\paragraph{\it Decaying Turbulence} 

If there is no turbulence production ($\mathcal{P}=0$), both the TKE and the associated dissipation rate will decay. In this case, the solution to Equations (\ref{eq:k_eq}) and (\ref{eq:epsilon_eq}) is \cite{bib27}:

\begin{align}
  k(t) & = k_0 \left( \frac{t}{t_0} \right)^{-m} \\
  \varepsilon(t) & = \varepsilon_0 \left( \frac{t}{t_0} \right)^{-(m+1)},
\end{align}

\noindent where:

\begin{equation}
t_0 = c_s m \frac{k_0}{\varepsilon_0}
\end{equation}

\noindent is the reference time ($k(t_0)=k_0$ and $\varepsilon(t_0)=\varepsilon_0$) and the decay exponent is:

\begin{equation}
m = \frac{1}{C_{\varepsilon 2}-1},
\end{equation}

\noindent which is equal to 1.09 for the standard value of $C_{\varepsilon 2}$. The exponent observed in measurements is somewhat higher, i.e., around 1.3 \cite{bib27}.

For decaying turbulence, the turbulence time scale can thus be written as:

\begin{equation}
\tau_L = \frac{k}{\varepsilon} = t \frac{k_0}{\varepsilon_0 t_0} = t \frac{1}{c_s m}
\end{equation}

\subsection{A Plasma Physics Analogy}

An analogy for the high Reynolds number transition is a controlled confinement transition in fusion plasmas \cite{bib40}. Here, a change in the topology of the magnetic field triggers a transition from “good” to “bad” confinement, where the temperature gradient decreases and the core turbulence level increases \cite{bib3}.

In a similar fashion, the mean velocity gradient decreases (mixing length increases) and the CL velocity fluctuations increase with increasing Reynolds number \cite{bib8}.

Thus, the low (high) Reynolds number range corresponds to good (bad) confinement, respectively. One physical aspect which may play a part is the emergence of large structures for the pipe flow case and correlated density--magnetic fluctuations for the fusion \mbox{plasma example.}

\subsection{Recommendations for CFD Practitioners}

In addition to being used for inlet boundary conditions in CFD simulations, the model output can be used as a complete, self-consistent model. These two applications are described in the following two sections.

For both applications, we recommend to use $\beta=1$ for the TKE, since this matches the standard turbulence model for low Reynolds numbers.

\subsubsection{Equilibrium Usage as Inlet Boundary Conditions for CFD Simulations}

The standard turbulence models in CFD simulations assume equilibrium flow; therefore we set the AA turbulence production-to-dissipation ratio to be equal to one. For this case, we compare the LIKE algorithm to our proposed equilibrium model in Table \ref{tab:LIKE_comp}. Some comments are in order regarding the contents of the table:

\begin{itemize}
  \item L: The expressions are taken from \cite{bib8}. Note that $\ell_{\rm LIKE}/\ell_{\rm AA} = 1/\kappa_g \sim 3$.
  \item I: The mix power law expression is derived in \cite{bib1} with more details provided in \cite{bib7}. The equilibrium model formulation is from Equation (\ref{eq:I_AA_output}).
  \item K: The mix and equilibrium expressions are from \cite{bib8} and Equation (\ref{eq:k_AA_def}).
  \item E: For the equilibrium model, we define $C_{\mu,{\rm AA}}^{3/4} = \left( B_g + \frac{3}{2} A_g - \frac{8 C_g}{3 \sqrt{Re_{\tau}}} \right)^{-3/2}$, see \mbox{Equation (\ref{eq:C_mu_AA}).} There is a difference of a factor $C_{\mu}^{1/4}$ between the TKE dissipation rates, which (partially) compensates for the difference between the length scales. Note that $C_{\mu,{\rm AA}}$ is a function of $Re_{\tau}$, whereas the LIKE algorithm uses the fixed standard value $C_{\mu,{\rm B}}$ \cite{bib8}.
\end{itemize}

We are of the opinion that consistent units must be used to define all turbulent quantities, i.e., either CL or AA. The mixed TI ($I_{\rm mix}$) is not straightforward to interpret and thus caution is advised for this definition.

The LIKE algorithm can be viewed as a CL model, since the used mean velocity $U_m$ cancels out for the TKE. However, $I_{\rm mix}$ remains a mixture of CL and AA quantities. The LIKE algorithm could be made into a consistent CL model by replacing the existing mixed TI definition with the CL TI definition presented in \cite{bib8}.

Thus, the AA-based equilibrium model we propose is more consistent than what is presented in the LIKE algorithm. Another advantage of the equilibrium model is that wall roughness is taken into account through the friction factor.

\begin{table}[!ht]
\caption{A comparison between the LIKE algorithm \protect\cite{bib32} and the proposed equilibrium model.} 
\centering 
\begin{tabularx}{\textwidth}{lll} %
\toprule 
\textbf{Source} & \textbf{LIKE Algorithm} & \textbf{Equilibrium Model} \\  
\midrule 
L   &  $\ell_{\rm LIKE} = \ell_{m,{\rm N,CL}} = 0.14 \times \delta$ & $ \ell_{\rm AA} = \langle \ell_{m,{\rm G-H}} \rangle_{\rm AA} = 0.14 \kappa_g \times \delta$ \\
I   &  $I_{\rm LIKE} = I_{\rm mix} = 0.16 \times Re_D^{-1/8}$ & $I_{\rm AA} = \sqrt{\left[ B_g + \frac{3}{2} A_g - \frac{8 C_g}{3 \sqrt{Re_{\tau}}} \right] \times \frac{\lambda}{8}}$ \ \\
K   & $k_{\rm LIKE} = k_{\rm mix} = U_m^2 I_{\rm mix}^2$  & $k_{\rm AA} = U_m^2 I_{\rm AA}^2$ \\
E   & $\varepsilon_{\rm LIKE} = C_{\mu,{\rm B}} \times \frac{k_{\rm LIKE}^{3/2}}{\ell_{\rm LIKE}}$  & $\varepsilon_{\rm AA} = C_{\mu,{\rm AA}}^{3/4} \times \frac{k_{\rm AA}^{3/2}}{\ell_{\rm AA}} = \frac{k_{\rm AA}^{3/2}}{L_{\rm AA}}$ \\
\bottomrule 
\end{tabularx}
\label{tab:LIKE_comp} 
\end{table}

\subsubsection{Non-Equilibrium Usage as a Standalone Model}

The recommendation for the use of the non-equilibrium model as a standalone model is to use the AA expressions in Section \ref{subsec:model_output} with $\beta=1$.

\subsection{Known Model Issues}

Our model is not able to capture low Reynolds number effects such as the decrease in $C_{\mu}$ \cite{bib19,bib23}. This is because the hyperbolic tangent functions used can only describe a single transition, which, in our case, is the high Reynolds number transition.

\section{Conclusions}
\label{sec:conc}

An algebraic mixing length non-equilibrium turbulence model has been developed to capture the high Reynolds number transition observed in pipe flow. The model equations have been derived to take the turbulence production-to-dissipation ratio explicitly into account. We provide area-averaged (integral) quantities and examples to match the Princeton Superpipe measurements used to calibrate the model, both for smooth and rough pipes.

The impact of isotropic or non-isotropic turbulence is investigated and {relevant figures-of-merit with area-averaged scaling are included}, 
such as turbulent viscosity, $C_{\mu}$ and time scales. We expect the predictions to be valid both for pipes and similar geometries such as closed (open) channel flow, see \cite{bib17} (\cite{bib39}), 
 respectively. It is essential to validate the model performance in the future; however, we do not know of either direct numerical simulations or alternative measurements for relevant friction Reynolds numbers.

A next step could be to use a similar non-equilibrium approach for more complex one- or two-equation turbulent viscosity models.

Future research could focus on generalising to rotating flows, see, e.g., an extended expression for $C_{\mu}$, which has been proposed \cite{bib26}.

\vspace{6pt}

Supplementary Materials: The following supporting information can be
downloaded from: \url{https://www.researchgate.net/publication/373108195\_Supplementary\_Information\_An\_algebraic\_non-equilibrium\_turbulence\_model\_of\_the\_high\_Reynolds\_number\_transition\_region}

\funding{This research received no external funding.}



\dataavailability{Data availability is not applicable to this article as no new data were created or analysed.}

\acknowledgments{We thank Alexander J. Smits 
 for making the Princeton Superpipe data publicly available.}

\conflictsofinterest{The author declares no conflicts of interest.}

\begin{adjustwidth}{-\extralength}{0cm}

\reftitle{References}

\PublishersNote{}
\end{adjustwidth}
\end{document}